\documentclass[runningheads]{llncs}
\usepackage{graphicx}
\usepackage{float}
\usepackage{tikz}
\usepackage{multirow}
\usepackage{subfigure}
\usepackage{xcolor}

\graphicspath{ {./figures/} }
\def\code#1{\texttt{#1}}

\begin{document}
\title{Multi-GPU Acceleration of the iPIC3D Implicit Particle-in-Cell Code}
\titlerunning{Multi-GPU Acceleration of the iPIC3D code}
%
\author{Chaitanya Prasad Sishtla\inst{1} \and Steven W. D. Chien\inst{1} \and Vyacheslav Olshevsky \inst{1} \and
Erwin Laure\inst{1} \and Stefano Markidis\inst{1}}
\authorrunning{C. P. Sishtla et al.}
%
\institute{KTH Royal Institute of Technology, Stockholm, Sweden \\
\email{\{sishtla,wdchien,slavik,erwinl,markidis\}@kth.se}}
\maketitle              
\begin{abstract}
iPIC3D is a widely used massively parallel Particle-in-Cell code for the simulation of space plasmas. However, its current implementation does not support execution on multiple GPUs. In this paper, we describe the porting of iPIC3D particle mover to GPUs and the optimization steps to increase the performance and parallel scaling on multiple GPUs. We analyze the strong scaling of the mover on two GPU clusters and evaluate its performance and acceleration. The optimized GPU version which uses pinned memory and asynchronous data prefetching outperform their corresponding CPU versions by $5-10\times$ on two different systems equipped with NVIDIA K80 and V100 GPUs.
	

\end{abstract}
\section{Introduction}
The advent of large supercomputers with multiple accelerators per computational node is impacting the development of large scientific applications. The two current largest supercomputers in November 2018 Top500 list, \emph{Summit} and \emph{Sierra}, feature six and four V100 NVIDIA GPUs per node respectively providing a theoretical peak performance of 750 and 500 Tops/s (operations in mixed precision) \cite{markidis2018nvidia} per node. Hence, it is important to exploit the computational power from GPUs on supercomputers. The PIC method is one of the main tools for the simulation of plasmas~\cite{birdsall2004plasma}. The method was initially developed in the late Fifties and early Sixties and then further improved by using more sophisticated numerical schemes, such as semi-implicit and fully-implicit schemes~\cite{markidis2011energy}, and combining fluid and kinetic equations for plasmas~\cite{markidis2018polypic}. 

iPIC3D has been designed for the kinetic simulation of space plasmas on large supercomputers \cite{markidis2010multi}. Its main application is the study of magnetic reconnection in Earth's magnetotail  \cite{peng2015energetic} and dayside magnetopause, kinetic turbulence \cite{olshevsky2013energetics} and interaction of solar wind with Earth's magnetosphere \cite{peng2015formation,peng2015kinetic} and comets~\cite{1901.09638}. It works as stand-alone code and as part of a multi-physics framework, called Space Weather Modeling Framework (SWMF), for the simulation for space weather \cite{chen2017global}. During the last decades the code has been improved by using advanced parallelization strategies, optimized I/O that have been developed during European-Commission funded EPiGRAM~\cite{markidis2016epigram} and SAGE projects \cite{narasimhamurthy2018sage}. It is written in C++ and uses MPI for internode communication. The code has shown scalability up to 80\% parallel efficiency on one million MPI processes \cite{markidis2016epigram}. However, iPIC3D does not currently support execution on GPUs which limits its usage on supercomputers with GPUs.  Since 2008 several studies focused on PIC code porting to efficiently use GPU systems. The first seminal work on this topic is by Stantchev et al.~\cite{stantchev2008fast}: it presents a PIC porting to GPUs focusing on the optimization of the interpolation step. Optimization of the data layout in PIC codes are presented in Refs. \cite{decyk2011adaptable,decyk2014particle}. However, all these previous works use simple formulation of PIC algorithm, and porting of a semi-implicit PIC method does not exist in the literature. Our work aims to fill this gap and present the porting of a semi-implicit PIC method to GPUs. In particular, we focus on describing the steps for porting the iPIC3D application on multi-GPU systems: Tegner and Kebnekaise~\cite{2019arXiv190304364C} featuring NVIDIA K80 and Tesla V100 GPUs.

\section{Methodology}
\label{methods}
The iPIC3D simulation is initialized first by setting particle positions and velocities and assigning electric and magnetic field values on grid points ({\bf Initialization}). After the initialization, a computational cycle is repeated until the end of simulation. Each computational cycle consists of three basic steps: {\bf Fields solver} - where the electric and magnetic fields are calculated from the semi-implicit formulation of Maxwell's equations on a grid by solving a linear system, {\bf Particle Mover} - where new particle positions and velocities are computed using the electric and magnetic field values on the grid points and interpolating them at the particle positions, and {\bf Moments Calculaton} - where particle moments of the distribution function, such as density, current and pressure are calculated on the grid by interpolation. 

We performed a dedicated profiling of the problem that is considered in this work, the so-called GEM (Geospace Environmental Modelling) challenge~\cite{birn2001geospace} as shown in Table~\ref{table:cpu-prof}. Particle mover is clearly the most expensive operation and so we focus on the porting and optimization of the iPIC3D particle mover while calculation of fields and moments still remain on CPU. Detailed descriptions about the particle mover can be found in~\cite{markidis2018polypic,markidis2011energy,1901.09638}. We present the performance of the multi-GPU iPIC3D by reporting the harmonic mean of our main figure of merit - Millions of Particles Advanced per second (MPA/s) obtained by dividing the total number of particles in the simulation by the average time spent in the particle mover per computational cycle. The standard deviation is plotted as an error bar and shows minimum variability between different simulation runs. 

Each numerical experiment has a computational grid consisting of $64\times64\times32$ cells, with $4$ particle species. We perform a maximum of $3$ predictor-corrector iterations in the mover. Unless otherwise is specified, $216$ particles per cell is used for each species and each simulation is repeated six times with the first being a warmup run. For simplicity, when we refer to one K80 GPU in subsequent text, we refer to one GK210 GPU engine.

\begin{table}[]
	\centering
	\caption{Percentage of execution time per cycle for the three PIC steps in a typical iPIC3D benchmark run on a CPU, varying the number of particles per cell (ppc).}
	\begin{tabular}{|c|c|c|c|c|c|}
		\hline
		\multirow{2}{*}{Part of Code} & \multicolumn{5}{c|}{\% Time Spent}              \\ \cline{2-6} 
		& 27ppc   & 64ppc   & 125ppc  & 216ppc  & 343ppc  \\ \hline
		Fields solver                 & 6.12  & 2.79  & 1.47  & 0.87  & 0.55  \\ \hline
		Particle mover                & 68.81 & 71.42 & 72.23 & 72.77 & 73.14 \\ \hline
		Moments calculation           & 25.04 & 25.76 & 26.31 & 26.35 & 26.29 \\ \hline
	\end{tabular}
\label{table:cpu-prof}
\end{table}

\subsection{Porting to Multi-GPU Systems}
We use NVIDA CUDA for porting iPIC3D to GPU and associate each MPI process with one GPU device which is allocated according to the rank of the MPI process. Since the particle mover is responsible for updating the new position of particles, it requires the following information: Grid (geometry) information about the particle's neighbour nodes which remains unchanged throughout the simulation, values of the electromagnetic field on these grid nodes which is updated every computational cycle, and current position and velocity of the particle to be updated. The number of particles in the simulation may vary due to open boundary conditions and injection of particles from the simulation boundary.

For each particle species, GPU kernel of the particle mover is launched such that each thread is responsible for updating one particle. We thus enable more particles per cell while having limited GPU memory. Double precision is used in the entire computation process. We improve the particle mover in three stages, each being an optimization of the former.

{\bf Simple Synchronous Implementation.}
We first allocate memory for grid and field data on both the host and device as the sizes are known beforehand. Second, the grid data is copied during the initialization, and it will remain on the device for the entire course of the simulation. We allocate all the remaining available device memory for particles. To avoid future resizing, we allocate the same amount of memory for the particle data on the host. In each cycle the field data is copied to device memory, the CUDA kernel is launched and the new particle positions and velocities are copied back to host memory. The above process is repeated for each particle species and is performed in a blocking, synchronous manner.

{\bf Host Memory Pinning.}
\label{host-memory-pinning}
CUDA performs Direct Memory Access (DMA) through PCI-E to move data between the device and host. However, since the operating system allocates virtual memory in a pageable fashion, data in this address space must first be copied to a staging area in memory before DMA can be performed. We implement host memory pinning by replacing the allocation of host memory for field and particles with \code{cudaMallocHost()} or \code{cudaHostAlloc()}, the APIs provided by CUDA to perform page-locked memory allocation. In this prototype allocation, we uniformly allocate 3GB as the maximum possible size for each particle species.

\begin{figure}[h]
	\centering
	\includegraphics[width=\textwidth]{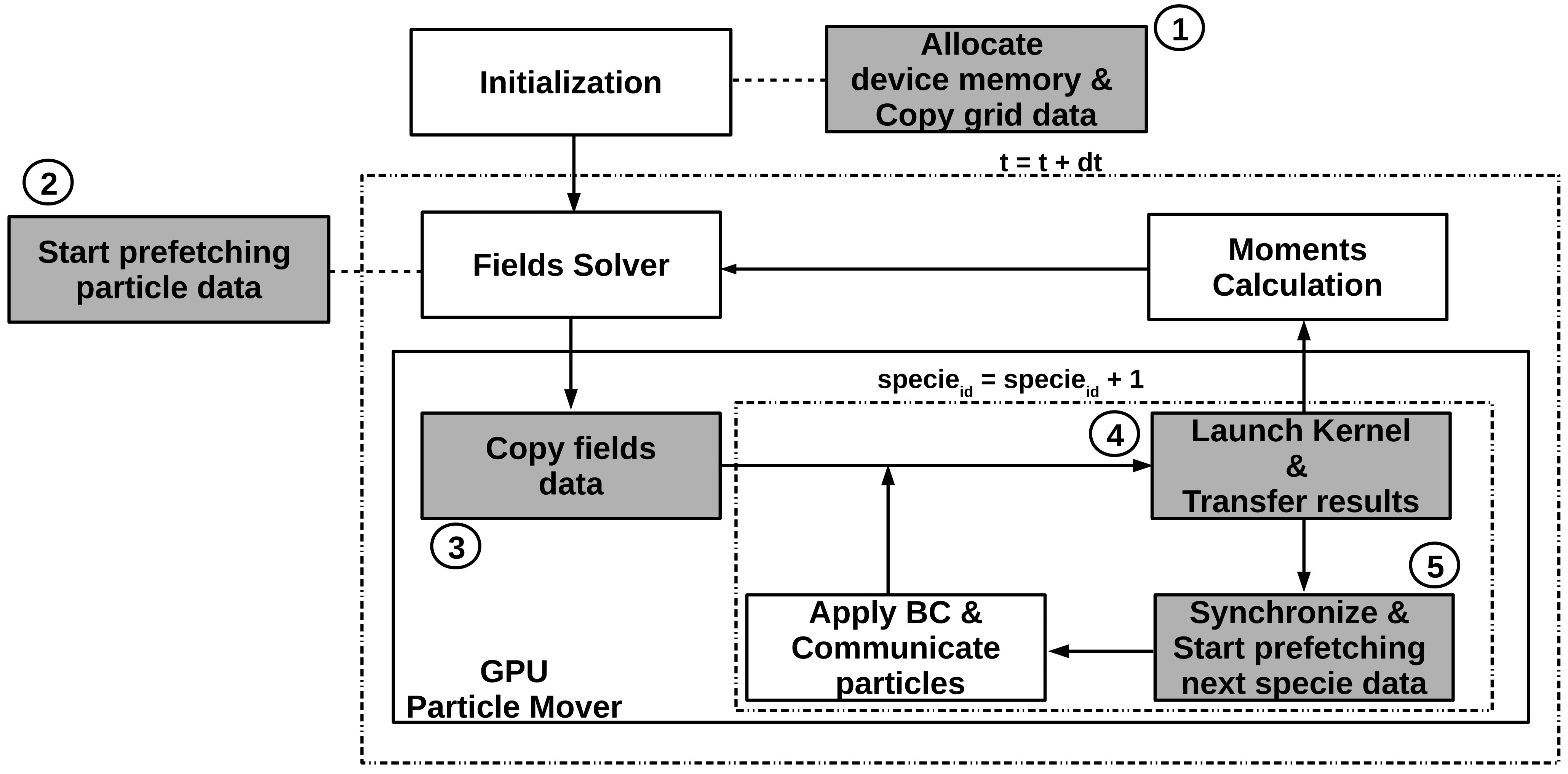}
	\caption{The flowchart of the iPIC3D code with the GPU particle mover using data prefetching is shown. The white blocks correspond to instructions executed on the CPU while the grey blocks correspond to CUDA code.  Dashed lines indicate where in the host the relevant CUDA code is called.}
	\label{data-flow-prefetch}
\end{figure}

{\bf Data Prefetching to GPU Memory.}
\label{data-prefetch}
The CUDA API supports asynchronous memory transfers by means of \code{cudaMemcpyAsync()} which can be effectively used to overlap CUDA calls with CPU computation. We use CUDA stream to ensure the order of kernel execution and data movement that are performed asynchronously to the host. The implementation of data prefetching is summarized in Figure~\ref{data-flow-prefetch}. Steps 1 and 3 are asynchronous data transfers which are respectively called before and after the fields solver. In Step 5 the mover kernel is followed by \code{cudaStreamSynchronize()}, and after synchronization, the updated data for the first species is copied to the host, and the asynchronous copying of the particle data of the next species to the GPU is initiated.

\section{Results}
 \label{results}
Even a simple synchronous (`naive') porting of the particle mover gives a substantial performance benefit on one GPU as demonstrated by the profiling results presented in Table~\ref{table:prof}. The acceleration, computed as the ratio of the mover execution times $A=T_\mathrm{CPU} / T_\mathrm{GPU}$, ranges from $4-4.5$ on K80 to $8.7$ on V100. Each improvement in the mover, use of pinned memory and prefetching, makes execution faster. The prefetch mover gives $30\%$ better acceleration than the simple mover on K80. The striking $A=25$ acceleration of the prefetch mover on V100 is affected by the slow execution on the corresponding CPU.

\begin{table}[]
	\centering
	\caption{The average time spent in the particle mover over $10$ cycles using a single GPU (one MPI process) in different testing environments.}
\begin{tabular}{|c|c|c|c|c|}
\hline
\multirow{2}{*}{Type of Node} & \multicolumn{4}{c|}{\begin{tabular}[c]{@{}c@{}}Particle mover execution times\\  (in seconds)\end{tabular}} \\ \cline{2-5} 
\multicolumn{1}{|l|}{}                              & CPU                       & Naive                   & Pinned                   & Prefetch                   \\ \hline
Tegner (Haswell+K80)                                        & $15.33$                    & $3.28$                   & $3.05$                    & $2.44$                      \\ \hline
Kebnekaise (Broadwell+K80)                                    & $15.20$                    & $3.84$                   & $3.44$                    & $2.87$                     \\ \hline
Kebnekaise (Skylake+V100)                                   & $36.82$                    & $4.20$                   & $2.02$                    & $1.43$                      \\ \hline
\end{tabular}
\label{table:prof}
\end{table}

\begin{figure}[h!]
	\centering
	\includegraphics[width=\textwidth]{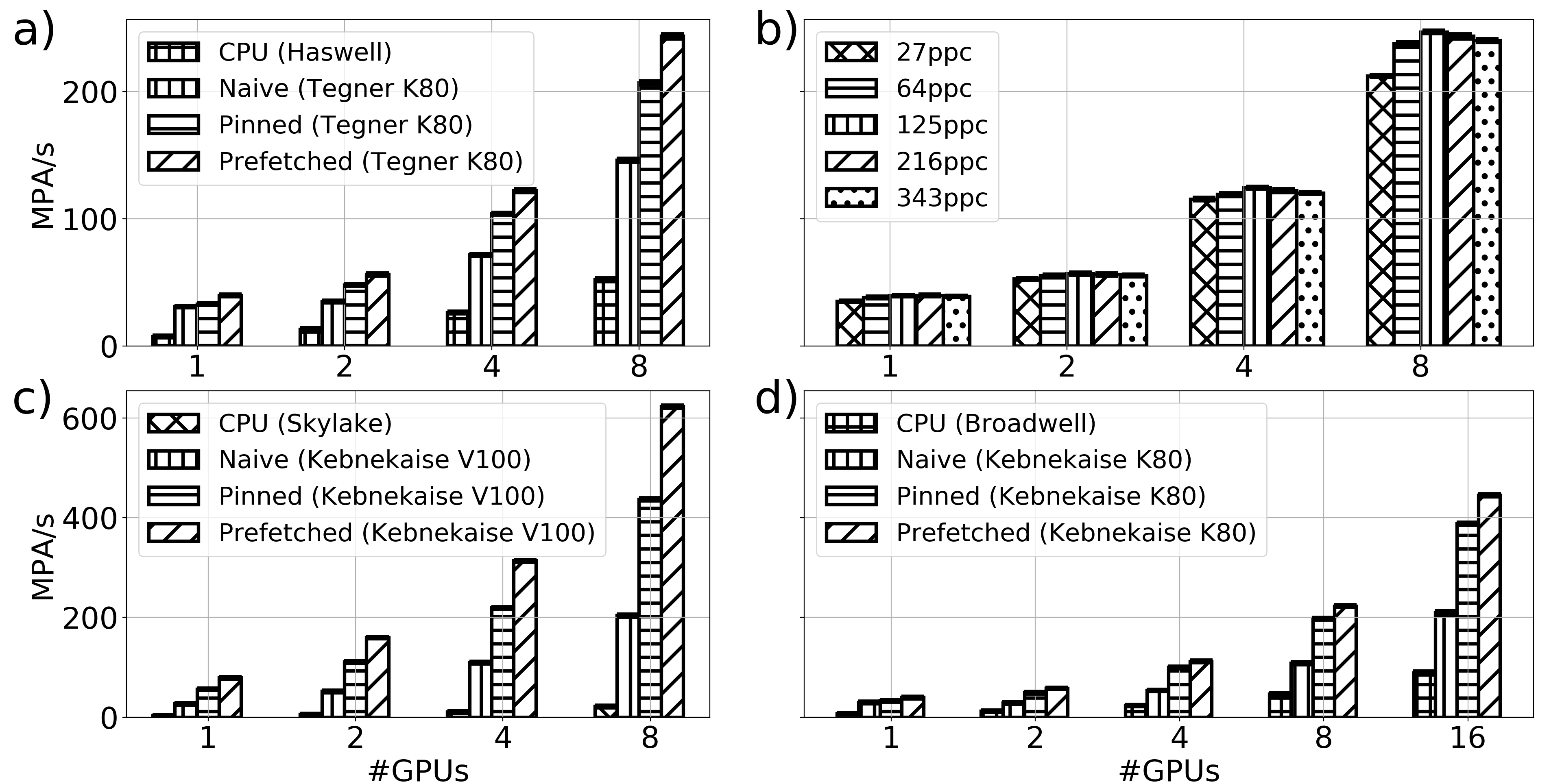}
	\caption{a) The performance of the GPU porting schemes compared on Tegner (using the Haswell CPU and the K80 node). b) The performance of the prefetched GPU porting schemes compared on the K80 nodes of Tegner by varying the number of particles per cell in the simulation. The performance of the GPU porting schemes compared for Kebnekaise. c) nodes with Broadwell CPUs and K80 GPUs; d) nodes with Skylake CPUs and V100 GPUs.}
	\label{all_results}
\end{figure}

To investigate the parallel performance of the three movers we did a strong scaling study. The same experiment (with the same initial configuration and number of particles) was repeated employing $2$, $4$, and $8$ MPI processes with $1$ GPU per MPI process on Tegner Haswell+K80 nodes. As a reference, a purely CPU study was also performed. 
The results of the scaling study for Tegner (Haswell+K80) are shown in Figure~\ref{all_results}a. The prefetch mover exhibits a peak performance (measured for $8$ GPUs) of $243$MPA/s, as compared to $206$MPA/s for pinned, $146$MPA/s for naive, and $52$MPA/s for the run at $8$ CPUs. Relative to the single-GPU run, the prefetch and pinned movers give the parallel speedup of $S=6.1$ at $N=8$ GPUs and the parallel efficiency of $E=S/N=76$\%. The speedup of the naive mover implementation only reaches $4.7$ giving $E=59$\%. The same parallel scaling experiments for up to $16$ MPI processes on Kebnekaise Broadwell+K80 nodes show similar results (Figure~\ref{all_results}d). The parallel speedup on $8$ nodes is similar to Tegner's K80 nodes, however the figures are somewhat lower, with $222$MPA/s given by the prefetch mover on $8$ GPUs. The parallel efficiency of the pinned and prefetch movers at $16$ GPUs is $73$\%, while for the naive mover it is only $44$\%. Therefore, asynchronous prefetching of particle data to GPUs is essential for the parallel performance of the mover. Scaling performance exhibited by the experiments on the Kebnekaise V100 nodes is shown in Figure~\ref{all_results}c. The results of the scaling study on V100 GPU show that the movers perform significantly better than on K80 both in terms of speedup and absolute figures. The prefetch mover performs the best with a peak performance of $622$MPA/s, as compared to $437$MPA/s for pinned and $204$MPA/s for the naive mover using 8 GPUs. We get a nearly optimal scaling for the prefetched mover which exhibits a parallel speedup of $S=7.9$ at $N=8$ GPUs resulting in a parallel efficiency of $98.8$\%. The parallel efficiency for the pinned and naive movers are $97.7$\% and $94.6$\% respectively. 


%

We ran a series of experiments on Tegner's Haswell+K80 nodes in the same setup as above, with prefetch mover, varying the number of particles in the system and the number of GPUs in order to study the impact of number of particles on parallel performance. The results are summarized in Figure~\ref{all_results}b. The parallel scaling appears very similar for different number of particles. However, there is a clear tendency of the simulation with $125$ particles/cell to outperform others. The peak performance for $125$ particles/cell at $8$ GPUs is $246$MPA/s, while for $27$ particles/cell it is lower, $212$MPA/s. The degrading in performance for the higher number of particles is not so significant, with $240$MPA/s for the run with $343$ particles/cell.



\section{Discussion and Conclusion}
\label{conclusion}
We have designed and implemented porting of the particle mover in the semi-implicit PIC code to GPUs . Numerical experiments using a typical space plasma physics simulations have shown that GPU movers clearly outperform the purely CPU implementation, being $5-10$ MPA/s faster. The experiments on K80 and V100 GPUs have shown that memory pinning and prefetching is essential to reach a good parallel performance. The best performance and scaling efficiency is exhibited by the prefetch mover. Its parallel efficiency reaches $73$\% on $16$ K80 GPUs, while the naive implementation of the mover results in the parallel efficiency of only $44$\%. Finally, the prefetch mover does not exhibit a substantial dependency of its performance and scaling on the number of particles in the simulation. The question, whether the performance of a GPU-ported particle mover depends on the number of particles, or some other parameters of the system, such as the number of predictor-corrector iterations, is worth further investigation. Further work should also include porting of the particle distribution function moment calculation to GPUs, or its possible merging with the mover phase of the computational cycle. Implementation of the efficient particle mover on GPUs is the most essential feature required for adapting iPIC3D to modern and forthcoming HPC architectures, and optimizing performance of the large-scale kinetic plasma simulations.  

\begin{flushleft}
	{\tiny This work has received funding from the European Commission H2020 program, Grant Agreement No. 801039 (EPiGRAM-HS). Experiments were performed on resources provided by the Swedish National Infrastructure for Computing (SNIC) at PDC Center for High Performance Computing and HPC2N. \par}
\end{flushleft}

\bibliographystyle{splncs04}
\bibliography{main}
\end{document}